\documentclass[11pt]{article}

\usepackage{graphics}   
\usepackage{endnotes}            
\usepackage{amsfonts}            
\usepackage{amssymb}             
\usepackage{amsthm}              
\usepackage{amsmath}            
\usepackage{algorithm}           
\usepackage{algorithmic}
\usepackage[letterpaper]{geometry}
\usepackage{rotating}            

\usepackage{setspace}                
\usepackage{natbib}
\bibpunct{(}{)}{;}{a}{,}{;}

\begin{document}

\title{
A Return to Biased Nets: New Specifications and Approximate Bayesian Inference\thanks{Paper prepared in memory of the late Thomas Fararo, whom the author thanks for his mentoring and input.  The author also thanks John Skvoretz for his helpful discussions.  This work was supported under NIH award 1R01GM144964-01.}
}

\author{
Carter T. Butts\thanks{Departments of Sociology, Statistics, Computer Science, and EECS, and Institute for Mathematical Behavioral Sciences; University of California, Irvine; \texttt{buttsc@uci.edu}}
}
\date{Published in the \emph{Journal of Mathematical Sociology}, DOI: 10.1080/0022250X.2024.2340137}
\maketitle

\begin{abstract}
The biased net paradigm was the first general and empirically tractable scheme for parameterizing complex patterns of dependence in networks, expressing deviations from uniform random graph structure in terms of latent ``bias events,'' whose realizations enhance reciprocity, transitivity, or other structural features.  Subsequent developments have introduced local specifications of biased nets, which reduce the need for approximations required in early specifications based on tracing processes.  Here, we show that while one such specification leads to inconsistencies, a closely related Markovian specification both evades these difficulties and can be extended to incorporate new types of effects.  We introduce the notion of inhibitory bias events, with satiation as an example, which are useful for avoiding degeneracies that can arise from closure bias terms.  Although our approach does not lead to a computable likelihood, we provide a strategy for approximate Bayesian inference using random forest prevision.  We demonstrate our approach on a network of friendship ties among college students, recapitulating a relationship between the sibling bias and tie strength posited in earlier work by Fararo.\\[5pt]
\end{abstract}

\theoremstyle{plain}                        
\newtheorem{axiom}{Axiom}
\newtheorem{lemma}{Lemma}
\newtheorem{theorem}{Theorem}
\newtheorem{corollary}{Corollary}

\theoremstyle{definition}                 
\newtheorem{definition}{Definition}
\newtheorem{hypothesis}{Hypothesis}
\newtheorem{conjecture}{Conjecture}
\newtheorem{example}{Example}

\theoremstyle{remark}                    
\newtheorem{remark}{Remark}


\section{Introduction}

Complex networks in social, biological, and physical systems continue to pose important challenges for modeling and analysis.  Among the most significant and long-standing is the theoretical problem of identifying mechanistically plausible generative processes that can describe networks in the amorphous, ``disordered but biased'' regime.  Such networks - which arise in settings ranging from friendship ties among high school students to protein aggregates - are broadly random (i.e., have high algorithmic complexity \citep{butts:sn:2001}), but show clear statistical patterns that strongly depart from the structure of homogeneous random graphs \citep[see e.g.,][among many others]{moreno.jennings:soc:1938,davis:asr:1970,przulj:bioinf:2007,felmlee.et.al:ans:2021}.  Many processes for generating such networks have been proposed (ranging from agent-based models \citep{hummon.fararo:sn:1995} to abstract mixing or rewiring processes \citep{watts:bk:1999} and potential-based models \citep{holland.leinhardt:jasa:1981}); of these, arguably the most useful have been frameworks that specify a wide range of biases, and that give rise to parametric specifications that are susceptible to inferential treatment.

The first fairly general framework of this type arose from work by the Rashevsky group at the University of Chicago in the late 1940s and early 1950s, under the rubric of ``random and biased nets'' \citep{solomonoff.rapoport:bmb:1951,rapoport:bmb:1951,rapoport:bmb:1957}.  Initially motivated by the problem of modeling the growth and structure of biological neural networks, the core ideas were quickly adapted and expanded by Anatol Rapoport and colleagues to study social systems \citep{rapoport:bmb:1953,rapoport:bmb:1953b}.  In its original incarnation, the biased net framework was defined in terms of a ``tracing process,'' by which (directed) ties from an initial node set were followed to their neighbors, with the process iterated outward until no new nodes are visited.  Ties emerging from a newly visited node could be directed back towards the ``parent'' who most recently nominated them, to a ``sibling'' nominated at the current remove, or to as-yet visited nodes; this logic was used in early work to specify approximate functions for the expected fraction of nodes reached at a given distance from a random starting point, in terms of the expected outdegree and the eponymous ``biases'' directing ties back to already-visited nodes.  This allowed the strength of parent, sibling, or other biases to be estimated from observed network data by tracing outgoing paths from randomly selected nodes and attempting to fit the parameters of the growth formula to the mean fraction of nodes reached at each distance (the \emph{structure statistics} of the graph).  

Among the most influential workers to take up the biased net approach was the late Thomas Fararo, who made numerous methodological and theoretical contributions to this research program.  His study \citep[with Morris Sunshine; ][]{fararo.sunshine:bk:1964} of a huge for its time high school friendship network of 417 individuals was a landmark demonstration of the approach, establishing its applicability to large systems in the field while also introducing methodological and theoretical advances (including innovative use of computer-aided analysis, improved tracing approximations, and the introduction of differential mixing biases).  He continued to expand on mixing biases in subsequent work, with particular eye towards connecting them with theories of stratification \citep{fararo:sn:1981,fararo.skvoretz:sn:1984,skvoretz.fararo:cpst:1986}, and also introduced the notion of closure biases that varied systematically with tie strength \citep{fararo:sn:1983}.  A number of his contributions were made with his student John Skvoretz, who subsequently made many significant contributions to the field \citep[e.g][]{skvoretz:sn:1985,skvoretz:sn:1990}.  Particularly notable is their joint work (with Filip Agneesens) on \emph{local} approximations to biased nets \citep{skvoretz.et.al:sn:2004}, a development on which this work builds.

Despite its long history and intuitive character, the biased net approach has in recent decades been overtaken by alternative frameworks for network modeling (most prominently the exponential family random graph models, or ERGMs \citep{lusher.et.al:bk:2012}).  A number of factors have led to this development.  One factor has been unclarity in the framework itself: the ``tracing process'' on which most work in the area was built was never a fully specified probability model, with numerous assumptions and approximations woven into it that prevented it from being easily developed into a complete theory.  A second (but related) factor was difficulty with inference.  Tracing-based analyses were not always definitive \citep[see e.g.][]{fararo.sunshine:bk:1964}, and were difficult to treat statistically.  Although efforts have been made to address these problems \citep{skvoretz:sn:1985,skvoretz:sn:1990,skvoretz.et.al:sn:2004}, a definitive solution remains an open problem.  

In this paper we return to this thread, developing a local version of the biased network framework introduced by \citet{butts:jms:2018} and inspired by the local approximations of \citet{skvoretz.et.al:sn:2004}.  We extend existing biased net specifications by introducing a new type of event - the inhibitory event - that provides a means of avoiding degeneracy arising from the sibling bias.  We also introduce a new and flexible approach to biased net inference based on approximate Bayesian computation, which combines ideas from machine learning and Bayes linear statistics to infer posterior quantities without explicit probability calculations.  We illustrate our approach by a proof-of-principle simulation study, and by an application to a university friendship network that empirically demonstrates the co-variation of the sibling bias with tie strength suggested by \citet{fararo:sn:1983}.

\section{Model Specification for Biased Nets}

In its pre-local specifications, the biased net process was described incompletely in terms of a tracing process; as we will be working here with more recent specifications, we provide only a brief summary sufficient to motivate the choice of terms and effects encountered later.  (See e.g. \citet{fararo.sunshine:bk:1964}, chapter 2 for a typical exposition.)  The tracing process begins with an initial seed node (or sometimes a set thereof) and follows their outgoing ties (often called ``axons'' or ``axones'' in early work, a holdover from the theory's origins in modeling neural systems); this constitutes the 0th generation of the trace.  The set of vertices out-adjacent to the 0th generation is then the 1st generation, and in general the set of vertices whose minimum geodesic seed-target distance is equal to $k$ comprise the $k$th generation.  The tracing process then deals with the destination of ties sent by the $k$th generation.  Given that some vertex at generation $k$ sends some number of outgoing ties (typically approximated as a fixed value), a biased net process postulates a series of unobserved events (``bias events'') that may affect their destinations.  For instance, the newly reached vertex may, with some probability $\pi$ (the ``parent bias'') send a tie back to the vertex that nominated them (i.e., a reciprocating tie).  Likewise, every vertex in generation $k$ with the same parent has an incoming shared partner; this is postulated to generate an event that with probability $\sigma$ (the ``sibling bias'') sends a tie from the focal vertex to its sibling (with one event per sibling).  One can further conjecture that the coincidence of an incoming edge from a sibling and the parental incoming shared partner provides another chance of prompting an edge to said sibling, an effect referred to as a ``double role bias'' ($\rho$).  Other types of biases can be created, e.g for specifically directing ties to particular groups \citep[see e.g.][]{fararo.sunshine:bk:1964,fararo:sn:1981,fararo.skvoretz:sn:1984,skvoretz.fararo:cpst:1986}.  The nominations arising from these bias events are posited to ``deduct'' ties from ego's outdegree, by sending ties ``inward'' to already reached vertices rather than ``outward'' to new ones.  With appropriate simplifying assumptions, this can be used to derive the expected number of new ``outward'' contacts for a newly reached vertex \citep{fararo.sunshine:bk:1964}, which can then be used to produce a difference equation describing the expected number of vertices reached at every step in the tracing process (given a random seed).  Fitting this growth model to empirically observed random traces allows one to fit (e.g., by least squares) the bias parameters, which are interpreted as the probabilities of the latent bias events.

Biased net theory was innovative in providing an intuitive way of describing non-random statistical structure in networks in terms of hidden events that provide a ``prompt'' or context for tie formation.  The ability of the theory to provide a direct (if approximate) link between these event rates and observed network structure (particularly in terms of quantities relevant to diffusion) was a major innovation.  However, there are, as noted, many loose ends in the basic theory.  In principle, events could compete with each other (in which case, the order of their resolution matters), but this does not seem to have been investigated in early work.  Nor are corner cases like exhaustion of ties considered.  Dynamics obtained are in expectation, and do not take into account the discreteness of the underlying state space.  \citet{skvoretz:sn:1985,skvoretz:sn:1990} provides ways of clarifying or rectifying some of these difficulties, though to date no fully developed version of the tracing model has been proposed.  Instead, there has been a shift in the literature towards local specifications, on which we build here.

\subsection{Conditional Specification}

Although still working within the tracing paradigm, \citet{skvoretz:sn:1985,skvoretz:sn:1990} derives approximations for some conditional tie probabilities (and the dyad census), a move towards a local view of biased net specification that was more completely realized by \citet{skvoretz.et.al:sn:2004}.  Skvoretz et al. obtain approximations to the conditional probability of an edge given the rest of the graph (the full conditionals), given a local view of the tracing process.  They also derive such conditionals (under similar conditions) for dyads and triads, though we focus here on the most basic case.  Expressing the conditional approximation in the notation of \citet{butts:jms:2018}, it can be given the general form
\begin{equation}
\ln \Pr\left(Y_{ij}=0|Y^c_{ij}=y^c_{ij},\theta^*,t\right) \approx t\left(i,j,y^c_{ij}\right)^T \log \theta^*, \label{eq_condnot}
\end{equation}
where, for graph support $\mathcal{Y}$ on vertex set $V$, $t:V,V,\mathcal{Y}\mapsto \{0,1,\dots\}^p$ is a vector of statistics, $\theta^* \in [0,1]^p$ is a parameter vector, and $Y^c_{ij}$ refers to all elements of the graph $Y$ except the $i,j$ edge variable.  Typically, $\mathcal{Y}$ is taken to be the set of all loopless digraphs on $V$, although this is not essential.  Here, the statistics in $t$ are understood to refer to counts of potential bias events affecting the $i,j$ edge (assumed to depend upon the rest of the graph structure), with $\theta^*$ being the associated conditional probabilities that the events \emph{fail} to be realized.  Although the above is mathematically direct, a more interpretively natural equivalent of Equation~\ref{eq_condnot} is
\begin{equation}
\Pr\left(Y_{ij}=1|Y^c_{ij}=y^c_{ij},\theta,t\right) \approx 1-\prod_{k=1}^p\left(1-\theta_k\right)^{t_k\left(i,j,y^c_{ij}\right)} \label{e_bn}
\end{equation}
where $\theta=1-\theta^*$.  This expression clarifies the logic of the local biased net model: each potential $i,j$ edge can be thought of as being exposed to a series of independent bias events, each of which has probability $\theta_k$ to be realized.  An edge then results if and only if at least one bias event ``fires.'' Including at least one constant within $t$ (i.e., $t(i,j,y^c_{ij})\propto 1$) provides a way to specify a base rate of tie formation, with events for parent, sibling, or other biases being easily added.  Specifically, we may represent the parent bias by the statistic
\[
t(i,j,y^c_{ij}) = (y^c_{ij})_{ji},
\]
the sibling bias by the statistic
\[
t(i,j,y^c_{ij}) = \sum_{k\in V\setminus\{i,j\}}(y^c_{ij})_{ki} (y^c_{ij})_{kj},
\]
and the double role bias by the statistic
\[
t(i,j,y^c_{ij}) = (y^c_{ij})_{ji} \sum_{k\in V\setminus\{i,j\}}(y^c_{ij})_{ki} (y^c_{ij})_{kj}.
\]

Although \citet{skvoretz.et.al:sn:2004} use tracing-process ideas in deriving their conditional specification, it should be noted that this scheme marks a conceptual shift from an emphasis on biases as factors that impact the rate at which new nodes are visited in a tracing process to bias events as generators of social ties (rather like the foci of \citet{feld:ajs:1981}).  It also represents a shift towards an explicitly probabilistic approach to inference: Skvoretz et al. propose to use the conditional specification to perform maximum pseudolikelihood inference (MPLE) for biased net parameters from observed data, in similar fashion to the use of MPLE for ERGMs \citep{strauss.ikeda:jasa:1990}.  This in principle leads to natural approximations to the standard error, among other useful quantities.

\subsubsection{The Conditional Specification is Ill-Posed}

Although the local specification is both general and intuitive, the approximations proposed in \citet{skvoretz.et.al:sn:2004} were not derived from a joint distribution for the full network $Y$.  This raises the possibility that there does not exist a joint distribution having Equation~\ref{e_bn} as full conditionals.  (This problem does not arise with ERGMs because they are specified in terms of their joint distribution - thus, the full conditionals obtained from an ERGM are always realizable.  Attempting to build models ``bottom up'' from proposed conditionals without a known joint distribution can often lead to difficulties, as noted by e.g. \citet{chen.ip:jscs:2014}).  If there is no random graph $Y$ whose full conditionals can be described by Equation~\ref{e_bn}, then the associated specification is ill-posed.  This turns out to be the case, at least for some some models.  Here we provide a proof of this result, by counterexample.

We begin by assuming an order-3 digraph with adjacency matrix $Y$ and vertex set $V=\{i,j,k\}$, and a realization $y$ with edge set $E=\{(i,j),(i,k),(j,k)\}$.  The model family we will consider is a biased net with a baseline effect and a sibling effect; we denote the corresponding parameters in ``formation'' form as $\theta=(d,\sigma)$.  Our development also makes use of the elementary observation that, for arbitrary random variables $A,B,C$ with respective realizations $a,b,c$,
\[
\frac{\Pr(A=a|B=b,C=c)}{\Pr(B=b|A=a,C=c)} = \frac{\Pr(A=a|C=c)}{\Pr(B=b|C=c)}.
\]
(This follows immediately from the definition of conditional probability: $\Pr(A=a,B=b|C=c)=\Pr(A=a|B=b,C=c)\Pr(B=b|C=c)=\Pr(B=b|A=a,C=c)\Pr(A=a|C=c)$.) 
 
Our proof focuses on edge variables $Y_{ik}$ and $Y_{jk}$: specifically, we will show that the conditional probabilities assumed by the conditional biased net specification have contradictory implications for $\Pr(Y_{jk}=1|Y^c_{ik,jk}=y^c_{ik,jk})$.  Since we will be holding all non-$(i,k),(j,k)$ edge variables constant throughout, we simplify notation by defining the event $z\equiv Y^c_{ik,jk}=y^c_{ik,jk}$ and working with $\Pr(\cdot|z)$.

To show the contradiction, we first derive some basic relationships between the conditional pmfs of $Y_{ik}$ and $Y_{jk}$ and their corresponding marginals (all given $z$).  Employing the conditional definition of the biased net model (and $y$), we have
\begin{align*}
\frac{\Pr(Y_{jk}=1|Y_{ik}=1,z)}{\Pr(Y_{ik}=1|Y_{jk}=1,z)} &= \frac{1-(1-d)(1-\sigma)}{d}\\
&=\frac{\Pr(Y_{jk}=1|z)}{\Pr(Y_{ik}=1|z)},
\end{align*}
this last following from the relationship between conditionals and marginals.  Similarly, we may consider the case in which $Y_{ik}=0$,
\begin{align*}
\frac{\Pr(Y_{jk}=1|Y_{ik}=0,z)}{\Pr(Y_{ik}=0|Y_{jk}=1,z)} &= \frac{d}{1-d}\\
&=\frac{\Pr(Y_{jk}=1|z)}{\Pr(Y_{ik}=0|z)},
\end{align*}
and the case in which $Y_{jk}=1$,
\begin{align*}
\frac{\Pr(Y_{ik}=1|Y_{jk}=0,z)}{\Pr(Y_{jk}=0|Y_{ik}=1,z)} &= \frac{d}{(1-d)(1-\sigma)}\\
&=\frac{\Pr(Y_{ik}=1|z)}{\Pr(Y_{jk}=0|z)}.
\end{align*}

From the first pair of relationships, we may derive $\Pr(Y_{jk}=1|z)$ as follows:
\begin{align*}
\Pr(Y_{jk}=1|z) &= \Pr(Y_{ik}=0|z) \frac{d}{1-d}\\
&= (1-\Pr(Y_{ik}=1|z)) \frac{d}{1-d}\\
&= \frac{d}{1-d}\left[1-\Pr(Y_{jk}=1|z)\frac{d}{1-(1-d)(1-\sigma)}\right]\\
&=\frac{d}{1-d}\left[1+\frac{d^2}{(1-d)(1-(1-d)(1-\sigma))}\right]^{-1}\\
&=\frac{d(d+\sigma-d\sigma)}{d+\sigma(1-d)^2}.
\end{align*}

This is not the only way to derive the marginal, however - we may also use the third relationship.  Starting with $\Pr(Y_{jk}=0|z),$ we obtain
\begin{align*}
\Pr(Y_{jk}=0|z) &= \Pr(Y_{ik}=1|z) \frac{(1-d)(1-\sigma)}{d}\\
&= \Pr(Y_{jk}=1|z) \frac{(1-d)(1-\sigma)}{1-(1-d)(1-\sigma)}.
\end{align*}
Noting that $\Pr(Y_{jk}=1|z)=1-\Pr(Y_{jk}=0|z)$, we can then solve for the former:
\begin{align*}
\Pr(Y_{jk}=1|z) &= \left[1+\frac{(1-d)(1-\sigma)}{1-(1-d)(1-\sigma)}\right]^{-1}\\
&=d+\sigma-d\sigma.
\end{align*}

This contradicts our earlier result, since $\tfrac{d(d+\sigma-d\sigma)}{d+\sigma(1-d)^2} \neq d+\sigma-d\sigma$ for arbitrary $d,\sigma$.  (For instance, $d=\sigma=0.5$ leads to a marginal probability of 0.6 in the first case, and 0.75 in the second.)  It follows that the conditional probabilities assumed by the biased net model are incompatible, and the model family is ill-posed. QED.

\emph{Remark:}  The essential problem with the conditional specification is that it considers only the direct impact of bias events on edge formation, leaving out the impact of edge presence/absence on our inference regarding the (hypothetical) event.  Thus, if the presence of an incoming two-star for $Y_{jk}$ makes the $Y_{jk}$ edge more likely, the observation of such an edge must also provide evidence in favor of the edges comprising the hypothetical two-star itself (since the $j,k$ edge would be present more often when the two-star is present than when it is absent).  Conditional probabilities are intrinsically ``inferential'' rather than ``mechanistic'' -- they in a (loose) sense incorporate not only the factors that give rise to a result, but also the consequences of that result (whence its state may be indirectly inferred).  The biased net mechanism is simple, but does not in general lead to similarly simple conditional edge probabilities; this is the inverse of the usual ERGM case, where a simple conditional specification hides a potentially very complex underlying mechanism.

\subsection{Markov Process Specification}

Although the above demonstrates that the conditionally specified biased nets are problematic, \citet{butts:jms:2018} proposed an alternative family that resolves the coherence problem while retaining a very similar intuition.  This proposal departs from past practice in the biased net literature by discarding the notion of a tracing process altogether, instead constructing an explicit dynamic process based on Markov chains whose behavior is driven by bias events.  In particular, let us consider a  dynamic process in which $Y$ evolves in discrete steps, $\ldots,Y^{(0)},Y^{(1)},\ldots$.  At each iteration, a randomly chosen $i,j$ edge state is updated, with an update either toggling the edge state (i.e., changing present to absent, or absent to present) or leaving it intact.  Let $(i,j)$ refer to the edge chosen for updating at an arbitrary time, $\ell$.  The next graph state is then given by
\begin{equation}
Y^{(\ell+1)}_{gh} = \begin{cases} 1 & (g,h)=(i,j) \ \mathrm{and}\ u^{(t)} < 1-\prod_{k=1}^p\left(1-\theta_k\right)^{t_k\left(i,j,\left(Y^{(\ell)}\right)^c_{ij}\right)} \\ 0 & (g,h)=(i,j) \ \mathrm{and}\ u^{(t)} \ge 1-\prod_{k=1}^p\left(1-\theta_k\right)^{t_k\left(i,j,\left(Y^{(\ell)}\right)^c_{ij}\right)} \\ Y_{gh}^{(\ell)} & (g,h) \neq (i,j)\end{cases} \label{e_pgibbs}
\end{equation}
where $\ldots,u^{(0)},u^{(1)},\ldots$ is a set of iid uniform deviates on the $[0,1]$ interval.  As the \emph{dynamic} or \emph{mechanistic} conditional probabilities of an edge update here follow Equation~\ref{e_bn}, we refer to this as a Skvoretz-Fararo biased net process (SFBN).  This process resembles a Gibbs sampler, and indeed will be a Gibbs sampler for the joint distribution associated with Equation~\ref{e_bn} where it exists; in general, however, it is a pseudo-Gibbs sampler \citep{chen.ip:jscs:2014}, converging under weak conditions on $t$ and $\theta$ to an equilibrium distribution that does not have Equation~\ref{e_bn} as full conditionals.

Instead of being viewed as a Gibbs sampler, however, the SFBN process can be seen as a simple model of social dynamics, in which (1) opportunities for edge state changes occur at random, (2) the current state of the network generates potential bias events impacting the focal relationship, and (3) the focal relationship is formed/remains present if and only if at least one of these events occurs.  The events remain independent Bernoulli trials, as in the conventional biased net picture, and existing effects behave essentially as defined in past work.  Importantly, however, taking the equilibrium of the SFBN as a generative model gives us a well-defined way to specify and simulate draws from the BN distribution.

As one would expect, generation of draws from an SFBN can be performed by simply iterating the update of Equation~\ref{e_pgibbs} (i.e., MCMC).  For some classes of specifications, \citet{butts:jms:2018} showed that exact draws from the target distribution can be obtained using a variant of Coupling from the Past \citep{propp.wilson:rsa:1996}.  Both algorithms are implemented within the \texttt{sna} package \citep{butts:jss:2008b} for the R statistical computing system.

\subsection{Satiation Events}

A fundamental feature of the standard biased net events when applied to the Markovian specification---including parent, sibling, and double-role biases---is that they can only enhance the formation of edges.  In the original formulation of biased net theory, edges formed through these events were modeled as being deducted from the outdegree of ego (which was not modeled explicitly), so this was not inherently problematic so long as ego's outdegree was taken to be sufficiently large.  For a generative formulation of biased net theory, however, this formation-only construction is problematic.  Although models based solely on edge-enhancing effects are easy to simulate, they can be subject to the same density-explosion route to degeneracy seen in cases like the edge-triangle ERGM \citep{haggstrom.jonasson:jap:1999}.  \citet{butts:jms:2018} introduces \emph{dichotomized} versions of the sibling and double role biases as a tool for combating this effect: in these specifications, only the first incoming shared partner produces a potential bias event, and no additional events are generated from the presence of additional shared partners.  The resulting sibling and double role statistics are then respectively
\[
t(i,j,y^c_{ij}) = \max_{k\in V\setminus\{i,j\}}(y^c_{ij})_{ki} (y^c_{ij})_{kj},
\]
and
\[
t(i,j,y^c_{ij}) = (y^c_{ij})_{ji} \max_{k\in V\setminus\{i,j\}}(y^c_{ij})_{ki} (y^c_{ij})_{kj}.
\]
\noindent While dichotomization is only currently used for sibling or double role effects, we here refer generically to models where the dichotomized versions of such terms are used as ``dichotomized'' models.  Although dichotomization can avoid the density explosion for these terms, one can imagine similar effects arising from the joint action of numerous different biases that would be less easily suppressed.  To obtain a more general solution, we must introduce a way to specify mechanisms that \emph{suppress} edge formation.

Given the origins of biased net theory in neuroscience, it seems apt to observe that the behavior of actual neural systems is based on a combination of excitory mechanisms (like those traditionally studied in biased net research) and inhibitory ones.  Here, we thus introduce the notion of an \emph{inhibitory event}, whose occurrence suppresses edge formation or retention.\footnote{Inhibitory and dichotomized events were first implemented in version 2.4 of the \texttt{sna} package, but the former have not previously been described in print.}  Further, we propose a class of such events that we refer to as \emph{satiation events,} that are generated by extant ties.  Each tie currently sent by ego has a chance to ``satiate'' him or her in the context of evaluating a potential tie to alter, thus inhibiting it.  This embodies the social logic that ties consume finite temporal and attentional resources \citep{mayhew.levinger:ajs:1976}, making additional nominations increasingly difficult to sustain as one's current portfolio of outgoing ties increases.  Unlike a hard degree constraint, however, satiation can have a variable impact both across individuals and over time (plausibly representing the unmodeled impact of idiosyncratic factors competing with the focal network for social resources).

The SFBN specification can be extended to incorporate inhibitory events as follows.  In the context of the more general biased net family, we say that an inhibitory event for an $i,j$ edge variable is one whose realization prevents edge formation - i.e., an $i,j$ edge forms if a formation event occurs and an inhibition event does \emph{not}.  In the context of the Markovian biased net framework, the corresponding family of updating rules for an $i,j$ edge is 
\begin{equation}
Y^{(\ell+1)}_{gh} = \begin{cases} 1 & (g,h)=(i,j) \ \mathrm{and}\ u^{(t)} < P\left(i,j,(Y^{(\ell)})^c_{ij},\theta,\phi\right) \\ 0 & (g,h)=(i,j) \ \mathrm{and}\ u^{(t)} \ge P\left(i,j,(Y^{(\ell)})^c_{ij},\theta,\phi\right) \\ Y_{gh}^{(\ell)} & (g,h) \neq (i,j)\end{cases} \label{e_pinhib}
\end{equation}
where 
\[
P(i,j,y^c_{ij},\theta,\phi) = \left[\prod_{k=1}^{p'}\left(1-\phi_k\right)^{w_k\left(i,j,y^c_{ij}\right)}\right]\left[1-\prod_{k=1}^p\left(1-\theta_k\right)^{t_k\left(i,j,y^c_{ij}\right)}\right]
\]
is the generalized updating probability from Equation~\ref{e_pgibbs} and $w$ is a vector of inhibitory event statistics (with parameters $\phi$).  $P$ is then interpretable as the probability that at least one formation event for the $i,j$ edge variable is realized, while all inhibiting events fail.

Although many types of inhibitory events are possible, perhaps the most natural is based on \emph{satiation}: sending ties becomes increasingly unfavorable with increasing outdegree, eventually blocking tie formation.  A very simple way of capturing this effect is via an inhibitory event whose precondition is the presence of an outgoing tie from ego---each time that ego adds an edge, ego satiates with fixed probability (and, if satiated, will not add additional edges).  Formally, this term takes the form
\[
w(i,j,y^c_{ij}) = \sum_{k \in V \setminus \{i,j\}} (y^c_{ij})_{ik},
\]
i.e., the number of ties sent by $i$ to other vertices.  In keeping with the tradition of assigning distinct letters to bias event parameters, we here use $\delta$ as a shorthand to refer to the satiation event probability.

This notion of satiation has several appealing properties.  In addition to being an intuitive implementation of the classic observation that tie formation and maintenance are costly \citep{mayhew.levinger:ajs:1976}, it clearly prevents runaway edge formation: in particular, the probability of adding a new outedge to a given vertex when this mechanism is active will fall exponentially in outdegree.  This trivially blocks the density explosion route to degeneracy.  Its implementation is also computationally facile (requiring only constant time updates to the outdegree), making it easily scalable for use in large systems.

\section{Approximate Bayesian Inference for Biased Nets}

Inference for the parameters of a SFBN process is complicated by the inability to directly specify the likelihood, $\Pr(Y=y|\theta,\phi)$.  Moreover, unlike the ERGM case, we do not know the sufficient statistics for SFBN distributions with the exception of trivial cases (e.g., the $d$-only, or $d,\pi$ families).  We can, however, simulate draws from $Y|\theta,\phi$.  This suggests the use of approximate Bayesian inference strategies \citep{rubin:as:1984,newton.raftery:jrssB:1994,sunnaker.et.al:plos:2013,blum.et.al:ss:2013}.  The particular approach taken here builds on strategies employed in Bayes linear statistics \citep{goldstein.woof:bk:2007} and kernel Bayes Rule inference \citep{fukumizu.et.al:jmlr:2013}, which are founded on a notion of Bayesian inference (introduced by \citet{definetti:bk:1974,definetti:bk:1975}) based on \emph{conditional expectations} rather than probabilities.  For instance, the posterior expectation of parameter $\psi=(\theta,\phi)$ given observation $Y$ with respect to prior $f$ corresponds to  $\mathbf{E}_f \psi | Y$, i.e., the conditional expectation (with respect to distribution $f$) of $\psi$ given $Y$.  Subject to regularity conditions, one can extend this to the posterior expectation of any function $g$, i.e. $\mathbf{E}_f g(\psi) | Y$.  Choosing $g(\psi)=\psi^k$ provides arbitrary posterior moments, from which central moments (e.g., the posterior variance) can be obtained.  Posterior distributions can be recovered by choosing indicator functions, i.e., $g(\psi) = I(\psi<k)$; the expectation of $g$ is then the probability that $\psi<k$.  Evaluation at multiple choices of $k$ then allows posterior quantiles (thence the CDF) to be determined.  

This approach to Bayesian inference (sometimes called \emph{prevision}) is sometimes employed in settings where it is difficult or undesirable to specify explicit probability distributions \citep{goldstein:jrssB:1981}, but it can also be motivated on computational grounds.  Looked at through a conventional lens, finding the conditional expectation is a \emph{regression problem}, with dependent variable $\psi$ and independent variable $Y$ drawn from population distribution determined by $f$.  Indeed, this problem reduces to linear regression in the simplest case, a fact that is heavily exploited in the area of Bayes linear statistics \citep{goldstein.woof:bk:2007}.  We could, moreover, imagine approximating the conditional expectation of $\psi$ in practice by the simple algorithm of (1) simulating draws of $\psi$ from the prior; (2) simulating draws of $Y|\psi$; (3) regressing $\psi$ on $Y$; and then (4) predicting $\psi$ from the regression model at the observed data.  Where the relationship between $Y$ and $\psi$ is highly nonlinear, this is unlikely to work well.  However, we are not obligated to use a linear model: any estimator of the conditional expectation (i.e., a least-squares learner) can be used to approximate $\mathbf{E}_f \psi | Y$.  \citet{fukumizu.et.al:jmlr:2013}, for instance, employ kernel learning algorithms for this purpose.  Here, we use random forests \citep{breiman:ml:2001} as our tool for learning the conditional expectation surface (leading to what we refer to here as \emph{random forest prevision}).  We describe this approach in detail below.

\subsection{Random Forest Prevision} \label{sec_rfprevision}

Random forest prevision proceeds in the basic manner described above, using a random forest (RF) as a nonparametric approximator of the conditional expectation.  Random forests are a natural choice for this problem, as they are flexible and consistent learners,\footnote{I.e., under weak regularity conditions, their prediction surface will converge to the conditional expectation in the limit of sample size.} straightforward to train, show robust performance, and can be efficiently adapted to quantile regression (an equivalent to the above-described method for obtaining posterior CDFs).  We learn each model parameter separately, approximating the graph state $Y$ by a vector of statistics, $s(Y)$.  (We discuss choice of statistics below.)  Our procedure is as follows:

\begin{enumerate}
\item Specify a prior distribution, $f$, on $\psi=(\theta,\phi)=(\pi,\sigma,\rho,d,\delta)$.
\item Draw iid $\psi^{(1)}, \ldots, \psi^{(m)} \sim f$.
\item For each $\psi^{(i)}$, draw $Y^{(i)} \sim \mathrm{SFBN}(\psi)$.
\item For each $Y^{(i)}$, compute statistics $S^{(i)} = s\left(Y^{(i)}\right)$.
\item For each parameter $\psi_i$, train a random forest predictor $\hat{\psi}_i(S)$ by regressing $\psi_i^{(\cdot)}$ on $S^{(\cdot)}$ with a least-squares loss.
\end{enumerate}

Given observation $Y_{obs}$, we then approximate the posterior expectation $\mathbf{E}_f \psi_i|Y_{obs}$ by $\hat{\psi}_i\left(s\left(Y_{obs}\right)\right)$.  Posterior variances (and hence standard deviations) can be approximated by repeating the above training process for $\psi^2$ (leading to the predictor $\hat{\psi^2}_i$, and then using the approximation $\mathrm{Var}_f \psi_i | Y_{obs} \approx \hat{\psi^2}_i\left(s\left(Y_{obs}\right)\right) - \hat{\psi}_i\left(s\left(Y_{obs}\right)\right)^2$.  Posterior intervals can be obtained in direct fashion using random forest quantile regression, exploiting the fact that $\Pr(\psi_i < q| Y_{obs}) = \mathbf{E}_f I(\psi_i < q) | Y_{obs}$; the procedure remains as above with the exception of the response variable.  

One convenient feature of the above scheme is that we can easily use even complex priors (e.g., slab-and-spike mixtures), so long as we can simulate draws from them: the prior tacitly affects the analysis via setting the baseline distribution of $\psi$, and thereby the distribution of the simulated data, $Y|\psi$.  For SFBN families, beta distributions and/or mixtures of beta distributions and degenerate distributions concentrated at 0 are natural and easily simulated choices.  Simulation of SFBN realizations can be implemented by straightforward Markov chain Monte Carlo, or by exact sampling when inhibition effects are not included \citep{butts:jms:2018}.  

\subsection{Choice of Statistics} \label{sec_stats}

The random forest prevision scheme of Section~\ref{sec_rfprevision} approximates $\mathbf{E}_f \psi|Y_{obs}$ in two ways: first, we rely on random forest models based on finite simulation samples to estimate the conditional expectation surface; and second, we proxy the graph realizations $Y$ by statistics $s(Y)$.  When $s$ is sufficient for $\psi$ (as in ERGM inference), no information is lost by the latter substitution.  In general, however, we do not know the sufficient statistics for SFBN families, and we must instead select on heuristic grounds a collection of functions that approximately capture the information regarding $\psi$ in the graph (as is typical in approximate Bayesian computation \citep{sunnaker.et.al:plos:2013,blum.et.al:ss:2013}).

Here, we employ several sets of terms, based on both traditional biased net practice and general considerations regarding indicators of graph structure; all are combined into a single vector, which serves as an input to the random forest.  Because random forests do well at handling high-dimensional, highly correlated inputs, and because some functional expressions of similar properties may have more direct relationship to certain parameters than others, we do not employ dimension reduction prior to use (though this may improve performance if using other learning algorithms).  We investigate the relative importance of our selected statistics in Section~\ref{sec_simexp} below. Our statistics consist of the following:

\paragraph{Basic Graph-Level Indices:}  Density and edgewise reciprocity are closely related to $d$ and $\pi$, and are included.  We also include transitivity, which has an obvious connection to $\sigma$ and $\rho$.

\paragraph{Triad Census Statistics:}  Although transitivity is a joint function of several triad frequencies, the individual triad census statistics may also be predictive in some cases; here, we include all 16 of the Holland-Leinhardt \citep{holland.leinhardt:ajs:1970} statistics, normalized by the total number of triads.

\paragraph{Degree Statistics:}  As $\delta$ acts to shape the upper tail of the degree distribution, we include the (normalized) mean square outdegree; we also include the normalized mean square indegree, cross-product between indegree and outdegree (correlation), and isolate fraction as additional sources of information.

\paragraph{Structure Statistics:}  Traditionally, the structure statistics (the fraction of vertices reachable at each remove from a randomly chosen starting node) were the original target of biased net researchers.  To ensure a consistent dimension across graph sizes (and to reduce redundancy), we use a compressed representation of the structure statistics rather than the statistics themselves.  Specifically, we fit a five-parameter logistic function by least squares to the exact structure statistics, $F(x)$, i.e.
\begin{equation*}
  F(x) =  \gamma_4 - (\gamma_4 - \gamma_1) / [1 + (x/\gamma_3)^{\gamma_2}]^{\gamma_5}
\end{equation*}
where $\gamma_1$ and $\gamma_4$ are the minimum and maximum values, $\gamma_2$ sets the initial steepness, $\gamma_3$ sets the x-axis scale (and hence the inflection point), and $\gamma_5$ controls asymmetry.  These five parameters are then used as statistics within our model.

\paragraph{Cohesion and Spectral Statistics:}  Beyond the triadic level, closure effects can also lead to the formation of larger cohesive structures.  We thus include several cohesion-related measures, specifically the density of the network formed by Simmelian ties \citep{krackhardt:rso:1999}, the mean (degree) $k$-core number, and the standard deviation of the core number.  We likewise include several spectral statistics related to concentration of structure into a small number of core-periphery structures.  We compute for each graph the singular value decomposition, and for the square roots of the singular values calculate (1) the ratios of the $i+1$th value to the $i$th (for $i\in 1,2,3$), and (2) the fraction of singular values with magnitude greater than $1/n$.

\subsection{Simulation Experiment} \label{sec_simexp}

To examine the ability of random forest prevision to infer biased net parameters under reasonable conditions, we perform a simulation experiment.  Specifically, we first simulate a sample for and then train a RF prevision model, and then assess its performance on a second synthetic data set with parameters selected to cover a plausible range of values.  We also examine variable importance for the prevision model, which provides additional insights into the statistics that appear to be contributing the most information to the prevision process.

\paragraph{Prior Specification:}  As a reasonable base prior that is likely to be broadly useful in a range of cases, we employ a slab-and-spike design that mixes positive mass at 0 (i.e., no bias effect) and relatively diffuse beta distributions.  The exception to this rule is the prior for the base formation event $d$, for which no spike is used (as $d=0$ makes the empty graph an absorbing state of the SFBN process).  Specifically, we take all parameters $\theta=(\pi,\sigma,\rho,d,\delta)$ to be \emph{a priori} independent, with $\theta_i \sim Z W$, where $Z \sim \mathrm{B}(0.5)$ for all parameters other than $d$ and $Z=1$ for $d$, and $W\sim \mathrm{Beta}(a_i,b_i)$, where $\mathrm{B}$ is the Bernoulli distribution and $\mathrm{Beta}$ is the beta distribution.  For all parameters other than $d$, we take $a_i=0.5,b_i=1.5$, leading to a diffuse distribution with some mass concentration near the origin (reflecting a higher \emph{a priori} probability for smaller bias effects, without damping out the potential for high-weight effects).  For $d$, we use $a_i=5\times 10/(N-1), b_i= 5 \times (1-10/(N-1))$, where $N=|V|$ is the graph size.  For moderate to large $N$, this prior places relatively high weight on baseline mean degrees in the 1-15 range, while being diffuse enough to allow for higher or lower values.  We note that, under this prior, we treat every bias as having \emph{a priori} even odds of being present or absent, making it relatively conservative for detection of effects.  Adjusting the Bernoulli parameter provides a simple way to tune this effect.

\paragraph{Prior Sample and RF Training:} To form the RF training sample, $5 \times 10^5$ iid prior parameter draws were taken (i.e., draws of $\theta$ from the above beta mixtures).  For each prior draw, one network of order $N=100$ was generated using dichotomized and one using undichotomized closure effects.  For each of these networks, summary statistics were defined per Section~\ref{sec_stats}, leading to two sets of network statistics (one for each model class).  Biased net simulation was performed using the \texttt{rgbn} function from the \texttt{sna} package \citep{butts:jss:2008b} within the \texttt{statnet} library of the \texttt{R} statistical computing system \citep{handcock.et.al:jss:2008,rteam:sw:2022}; standard MCMC was used, with one independent chain per draw and $500 N^2 = 5 \times 10^6$ burn-in iterations discarded prior to selecting the final state.  Network statistics were calculated using the \texttt{sna} and \texttt{network} \citep{butts:jss:2008a} libraries.

Given the respective prior parameter and network samples, random forests were trained for the posterior expectation, mean square, and quantiles as described in Section~\ref{sec_rfprevision}.  Random forest training was performed using the \texttt{ranger} package \citep{wright.ziegler:jss:2017} with default settings (500 trees, mtry equal to the square root of the number of predictors, minimum node size 10, unlimited tree depth, splitting based on estimated response variance).  (As gains from hyperparameter tuning for random forests are usually minimal, we do not employ it here.)  The respective RF prevision models were then retained for subsequent prediction and analysis.

\paragraph{Target Sample and Prediction:} To assess inferential performance, we create a full factorial design with parameter values $\pi \in \{0,0.25,0.5,0.75\}$, $\sigma \in\{0, 0.1, 0.2, 0.3\}$, $\rho\in\{0,0.25,0.5\}$, $d \in \{1/(N-1), 3/(N-1), 6/(N-1)\}$, and $\delta \in \{0, 1/10, 1/5\}$, with 50 replications per condition (for a total of 21,600 replications on 432 unique parameter vectors; $N=100$).  For each replicate, one graph was drawn from the corresponding dichotomized model, and one from the undichotomized model, using the above simulation protocol.  Statistics were computed for each simulated network as per the above.  For each vector of statistics in each target sample, the corresponding RF prevision models were used to compute the posterior means and 95\% posterior intervals.  These were used to assess inferential performance as described below.

\paragraph{Inferential Performance:} Although our goal is to perform Bayesian inference, it can be useful for technique assessment to consider the frequentist properties of the resulting estimates.  Table~\ref{tab_sim} shows the bias and median absolute error (MAE) for the posterior mean as an estimator of the target parameter values, as well as the coverage of the 95\% posterior intervals, by parameter and dichotomization.  We note that bias is fairly small, and generally in the direction that is expected given the prior (which, for non-$d$ parameters, puts a large net weight on small parameter values).  MAEs are also small on the whole, though notably higher for the double-role bias ($\rho$).  This appears to reflect the fact that $\rho$ is generally hard to estimate: because the opportunity for double-role events only arises when one has an edge variable with both a reciprocating tie and at least one incoming shared partner, there are many fewer occasions to observe their consequences (particularly in sparse graphs).  The parent bias, $\pi$, while much better estimated, also shows somewhat higher error rates than other parameters for plausibly similar reasons.

\begin{table}[ht]
\centering
\begin{tabular}{rrrrrrrr}
  \hline\hline
 & \multicolumn{3}{c}{Undichotomized} & & \multicolumn{3}{c}{Dichotomized} \\
 & Bias & MAE & Coverage & & Bias & MAE & Coverage \\ 
  \hline
  $\pi$ & -0.049 & 0.085 & 0.998 & & -0.088 & 0.108 & 0.999 \\ 
  $\sigma$ & -0.023 & 0.058 & 0.980 & & 0.037 & 0.046 & 0.992 \\ 
  $\rho$ & -0.054 & 0.144 & 0.994 & & -0.072 & 0.158 & 0.994 \\ 
  $d$ & 0.032 & 0.007 & 0.979 & & 0.030 & 0.009 & 0.994 \\ 
  $\delta$ & -0.009 & 0.029 & 0.997 & & -0.005 & 0.023 & 0.999 \\ 
   \hline\hline
\end{tabular}
\caption{\label{tab_sim} Frequentist properties of the approximate Bayes estimators; coverage calculated for 95\% posterior intervals.}
\end{table}

Although posterior intervals are not confidence intervals (and intervals accurately covering $X\%$ of the posterior density need not have $X\%$ frequentist coverage), the coverage of the former often converges to the latter as sample size increases \citep{hartigan:jrssB:1966}.  It is unclear whether a similar phenomenon will occur for biased net models, given the nature of the dependence involved.  However, we may still examine the coverage properties of the posterior intervals for our test case, to gain a sense of what can be expected in practice.  Table~\ref{tab_sim} shows that the 95\% posterior intervals in the present setting are quite conservative from a frequentist standpoint, typically having coverage levels in excess of 99\%.  On average, then, the posterior interval nearly always covers the true parameter value, but the intervals are wider than would be needed for 95\% coverage.  (Of course, the standard Bayesian interpretation of the posterior interval holds regardless.)

\begin{figure}
\begin{center}
\includegraphics[width=\textwidth]{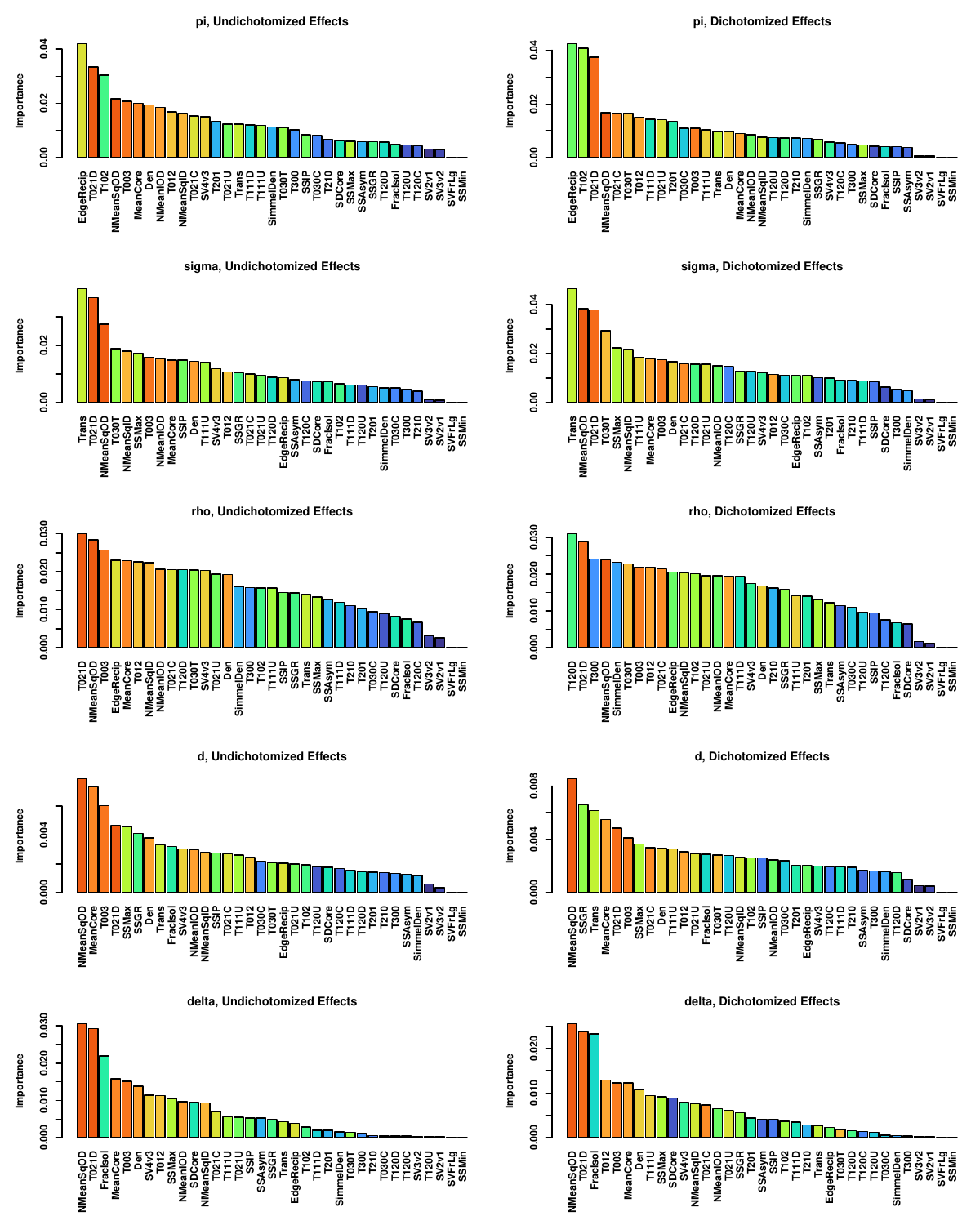}
\caption{\label{fig_varimp} Variable importance, by parameter, for $\mathbf{E}_f \theta|Y_{obs}$.  Colors indicate overall importance rank across cases (orange - high to dark blue - low).}
\end{center}
\end{figure}

\paragraph{Variable Importance:} Given that we currently lack formal results to guide selection of statistics for inference, it is useful to consider the extent to which those used here systematically contribute to posterior inference.  Here, we assess this via the \texttt{ranger} permutation-based variable importance procedure \citep{wright.ziegler:jss:2017,nicodemus.et.al:bioinf:2010}, which provides an index of the extent to which each statistic contributes to the reconstruction of $\theta$ given $Y$.  Figure~\ref{fig_varimp} shows importance scores for each statistic on each parameter, in the dichotomized and undichotomized cases.  Bars for each score are colored based on the mean rank of the statistic over all models (ranging from orange for consistently high importance statistics to blue for consistently low importance statistics).  Overall, the four most consistently important (highest mean rank) statistics for both model types are the normalized squared out-degree (NMeanSqOD), the normalized fractions of empty (T003) and open two-out triads (T021D), and mean core number (MeanCore).  Although relative importance varies somewhat between model types, the normalized mean squared indegree (NMeanSqID - more important for undichotomized models), density (Den), lone asymmetric dyad (T012), and transitive asymmetric triads (T030T - more important for dichotomized models) are also major contributors.  Contrary to expectations, the summarized structure statistics are not consistently strong predictors.  The minimum structure statistic term (SSMin) is the lowest ranked, but this is artifactual since it by construction takes the same value in the setting considered here.  More meaningfully weak predictors include the fraction of large singular values (SVFrLg), the relative sizes of the second and third singular values (SV2v1 and SV3v2, indicators of subsidiary cores), and the 120C cyclic triad (T120C).  It should be noted, however, that some statistics that are not consistently important are nevertheless important for some parameters.  For instance, edgewise reciprocity (EdgeRecip) is the most important statistic for recovering the parent bias in both models, and transitivity (Trans) is likewise the most important statistic for recovering the sibling bias.  The lone mutual dyad (T102) is another example, being important for the parent bias but little else, and the isolate fraction (FracIsol) is vital only for satiety.  Specialized statistics thus play a critical role.  We also note that no statistic used here is very close to sufficiency (which would be observable by a single statistic having extremely high performance, and others having none), and some parameters (e.g., the double-role bias) have high weights on a very large number of parameters.  This suggests the potential for development of new statistics with improved performance for biased net models, a theme to which we return in Section~\ref{sec_altlearn}.

\section{Application to the Webster Friendship Network}

\begin{figure}
\begin{center}
\includegraphics[width=0.5\textwidth]{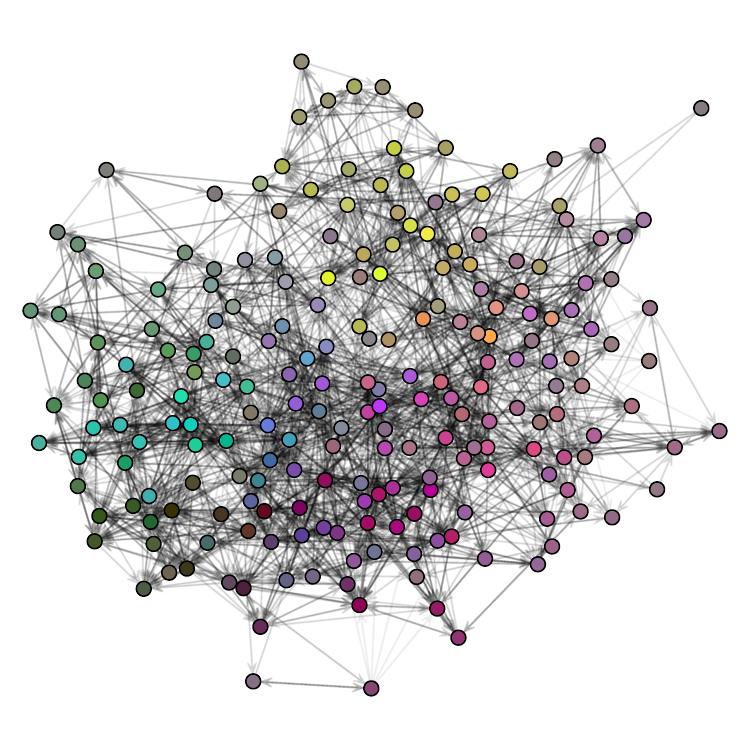}
\caption{\label{fig_netplot} Webster friendship attribution network.  Edges shaded by strength (darker=stronger); vertices colored by position on eigenvectors 2-4 of the graph adjacency matrix (more similar colors indicate greater similarity in subgroup membership).}
\end{center}
\end{figure}

\citet{freeman.et.al:sn:1998} reports a study conducted by Cynthia Webster of friendship attribution among students living in a residence hall at Australian National University ($N=217$).  Ties were elicited by a roster-supplemented name generator, followed by elicitation of perceived tie strength on a five-level scale.  We here interpret the relation as representing ego's attribution of friendship towards alter, and hence view it as directed (see Figure~\ref{fig_netplot}.  Although Freeman et al. do not provide the instrument employed, they indicate that the three highest ratings (which were also the most heavily used) were respectively described to subjects as ``Friend,'' ``Close Friend,'' and ``Best Friend.'' Because relations like ``Close Friend'' and ``Best Friend'' are recognizable labels that carry distinct cultural meaning (i.e., there exist norms within particular milieu regarding what counts as a ``Best Friend'' or other such tie, unlike numerical scores that are purely subjective), it is reasonable to view the data as a Guttman-like structure of hierarchically nested friendly relations, where each has meaning to the subjects but the stronger are constrained to entail the weaker.  Our analysis pursues this interpretation.

We here use random forest prevision to perform a biased net analysis of the Webster network, estimating parameters for each level of the relation (using the Guttman-like interpretation above); this allows us to not only to estimate effects for the network as a whole, but also allows us to see how biases change as one considers stronger and more intimate relations.  This approach echoes that used in early biased net research \citep[e.g][]{rapoport:bmb:1957} using ranked tie data, in which tracing was performed sequentially on the network formed by respondents' $k$th choice.

\subsection{Procedure}

The Webster network was thresholded at each measured level of strength, yielding five networks; estimates were obtained for each separately.

We use the same slab-and-spike priors for this analysis as in Section~\ref{sec_simexp}.  $5 \times 10^5$ iid prior draws were obtained for the RF training sample.  For each prior draw, one network of order matching the observed data size ($N=217$) was generated from the dichotomized and undichotomized models (respectively), and summary statistics were calculated for each as described in Section~\ref{sec_stats}.  Biased net simulation was performed using the \texttt{rgbn} function from the \texttt{sna} package \citep{butts:jss:2008b} within the \texttt{statnet} library of the \texttt{R} statistical computing system \citep{handcock.et.al:jss:2008,rteam:sw:2022}; standard MCMC was used, with one independent chain per draw and $500 N^2 \approx 2.4 \times 10^7$ burn-in iterations discarded prior to selecting the final state.  Network statistics were calculated using the \texttt{sna} and \texttt{network} \citep{butts:jss:2008a,butts:jss:2008b} libraries.

Given the respective prior parameter and network samples, random forests were trained for the posterior expectation, mean square, and quantiles as described in Section~\ref{sec_rfprevision}.  Random forest training was performed using the \texttt{ranger} package \citep{wright.ziegler:jss:2017} with default settings (500 trees, mtry equal to the square root of the number of predictors, minimum node size 10, unlimited tree depth, splitting based on estimated response variance).  As gains from hyperparameter tuning for random forests are usually minimal, we do not employ it here.  For the model selection problem, a (classification) random forest was trained on the combined biased net samples (for a total of $1 \times 10^6$ data points), with the outcome being the type of model (dichotomized versus undichotomized).  Posterior quantities were then approximated by prediction from the fitted random forests on the statistics from each of the five observed networks.

\subsection{Results}

\paragraph{Undichotomized Versus Dichotomized Effects:}  We begin by considering whether there is clear evidence for truncation in the impact of closure events (sibling and double role biases).  Although there is very weak evidence against dichotomization, the evidence is not conclusive: the estimated posterior probabilities for each network to be produced by the undichotomized model (versus the dichotomized counterpart) are respectively 0.55, 0.55, 0.56, 0.55, and 0.51 (low to high tie strength).  We can, however, exclude the possibility of strong evidence \emph{against} undichotomized effects (as might have been expected from the simulation study of \citet{butts:jms:2018}).  Given that both types of models are \emph{a posteriori} plausible, we consider posteriors for both cases in our subsequent analysis.

\begin{figure}
\begin{center}
\includegraphics[width=\textwidth]{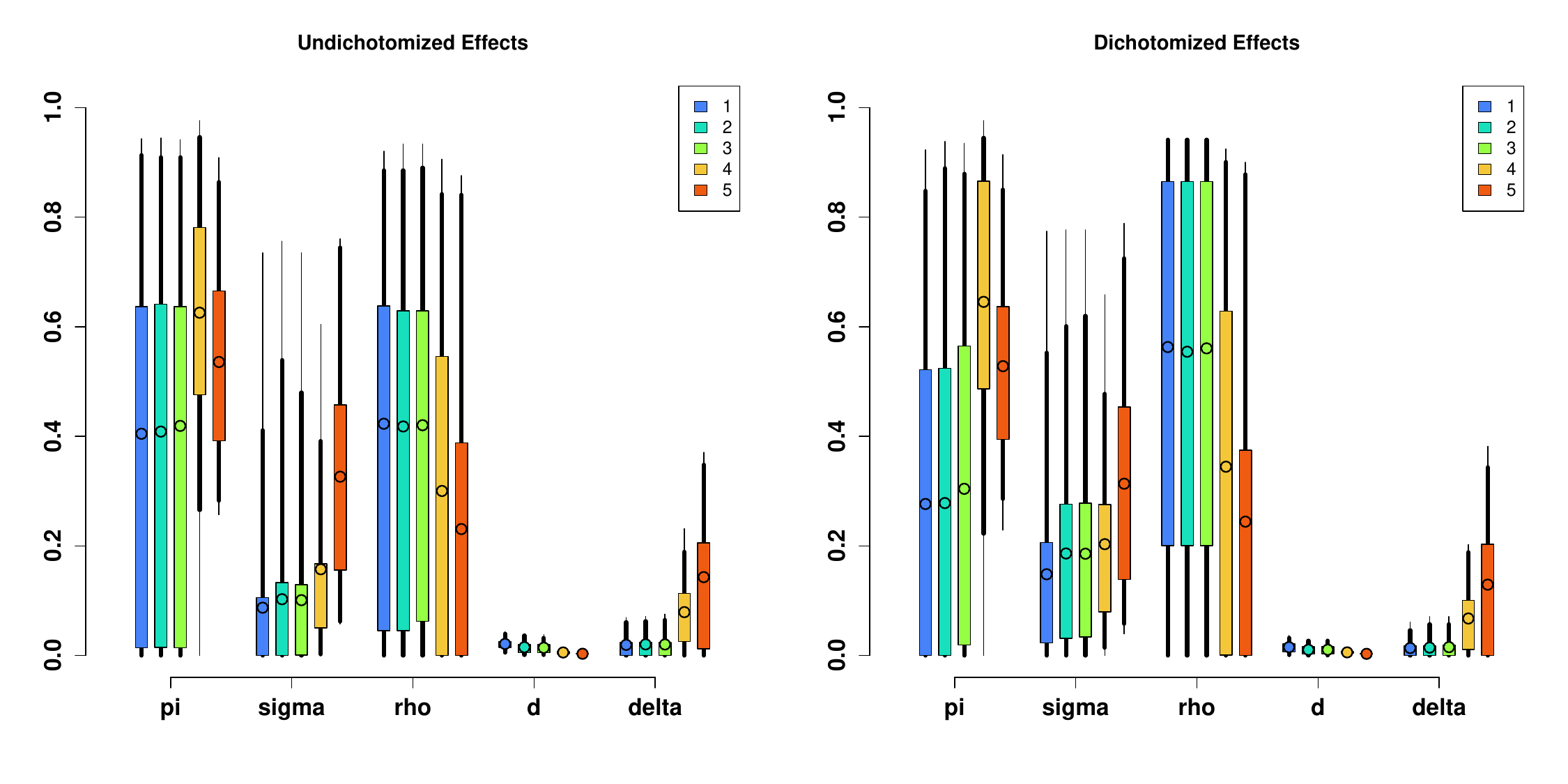}
\caption{\label{fig_effplot} Posterior estimates for Webster networks, by tie strength (color) and dichotomization.  Thin vertical lines show 95\% probability intervals, heavy lines show 90\% intervals, and colored boxes show the interquartile range; posterior means shown with highlighted points. Overall, stronger ties show greater reciprocity and closure (though not double role effects), along with lower base density and substantially greater inhibition.}
\end{center}
\end{figure}

\paragraph{Estimated Structural Biases:} Table~\ref{tab_coef} provides posterior expectations and standard deviations for each parameter, by tie strength and dichotomization.  Posterior intervals are depicted graphically in Figure~\ref{fig_effplot}.  Although there are small quantitative differences between the two types of effects, the posterior estimates are remarkably similar, and support the same interpretation.  Looking across different levels of tie strength, we see an average decline in $d$, as would be expected from moving from weaker to stronger (and, indeed, nested) relations.  This is not the only difference, however.  The parent bias, $\pi$, increases for strong friendships (suggesting enhanced tendencies towards reciprocation), as does the sibling bias ($\sigma$).  Being viewed as a friend by a common alter has a relatively modest effect on nominations for weak friendships, but becomes strong for best friends: in the dichotomized model, the presence of such an incoming shared partner configuration will induce an otherwise non-existent edge approximately 15\% of the time for weak friendships (discounting double-role effects), more than doubling to 31\% for best friends.  By contrast, we actually see a decline in double role bias for strong ties.  These two effects approximately cancel, meaning that the co-presence of an incoming tie and an incoming shared partner have roughly the same impact on friendship nominations for strong versus weak ties.  Since both parent and sibling biases are much weaker for weak ties, however, the \emph{sole} presence of an incoming edge or an incoming shared partner is less likely to induce nomination in this case: for strong ties, either circumstance is enough to generate a high level of nomination, but for weak ties the two must co-occur to have the same effect.

\begin{table}[ht]
\centering
\resizebox{\columnwidth}{!}{%
\begin{tabular}{rrrrrrrrrrrrr}
  \hline\hline
 &          & \multicolumn{2}{c}{$\pi$} & \multicolumn{2}{c}{$\sigma$} & \multicolumn{2}{c}{$\rho$} & \multicolumn{2}{c}{$d$} & \multicolumn{2}{c}{$\delta$}\\
 & Strength & $\mathbf{E} \pi$ & $\mathbf{sd(\pi)}$ & $\mathbf{E} \sigma$ & $\mathbf{sd(\sigma)}$ & $\mathbf{E} \rho$ & $\mathbf{sd(\rho)}$ & $\mathbf{E} d$ & $\mathbf{sd(d)}$ & $\mathbf{E} \delta$ & $\mathbf{sd(\delta)}$ \\
  \hline
Undichotomized & 1 & 0.405 & 0.387 & 0.087 & 0.229 & 0.423 & 0.416 & 0.021 & 0.013 & 0.019 & 0.040 \\
   & 2 & 0.409 & 0.389 & 0.103 & 0.248 & 0.418 & 0.417 & 0.015 & 0.013 & 0.020 & 0.045 \\
   & 3 & 0.419 & 0.373 & 0.101 & 0.239 & 0.420 & 0.414 & 0.014 & 0.011 & 0.020 & 0.048 \\
   & 4 & 0.626 & 0.344 & 0.157 & 0.207 & 0.300 & 0.410 & 0.005 & 0.024 & 0.079 & 0.093 \\
   & 5 & 0.536 & 0.167 & 0.326 & 0.293 & 0.231 & 0.293 & 0.003 & 0.002 & 0.143 & 0.157 \\
   &       &       &       &       &       &       &       &       &       &       &\\
Dichotomized & 1 & 0.276 & 0.377 & 0.148 & 0.207 & 0.563 & 0.361 & 0.015 & 0.025 & 0.014 & 0.033 \\
   & 2 & 0.278 & 0.395 & 0.186 & 0.192 & 0.554 & 0.360 & 0.010 & 0.024 & 0.014 & 0.035 \\
   & 3 & 0.304 & 0.374 & 0.186 & 0.181 & 0.561 & 0.368 & 0.010 & 0.022 & 0.015 & 0.037 \\
   & 4 & 0.645 & 0.355 & 0.203 & 0.227 & 0.345 & 0.389 & 0.006 & 0.005 & 0.067 & 0.086 \\
   & 5 & 0.528 & 0.129 & 0.314 & 0.233 & 0.244 & 0.306 & 0.003 & 0.018 & 0.129 & 0.142 \\
   \hline\hline
\end{tabular}
}
\caption{\label{tab_coef} Posterior means and standard deviations for the Webster network models.}
\end{table}

Tie strength also has an considerable impact on satiation.  Satiation effects are negligible at low strength, increasing sharply for close and best friends.  This occurs despite the simultaneous reduction in the baseline tie formation rate, $d$, and indeed the baseline formation is far below the level at which satiation would have a noticeable effect.  This increase, however, coincides with increases in the sibling bias, an effect that can if unchecked induce a density explosion.  Low baseline tie formation is not sufficient to prevent runaway tie formation without inducing excessive sparsity, but satiation can do so by selectively damping out ties from high-outdegree actors. 

Putting the pieces together, our models can be mechanistically interpreted as follows.  In weak friendship relations, baseline density is relatively high, with some tendencies towards reciprocity and ISP-induced closure; while relatively weak on their own, they become much stronger when combined.  Satiation plays little role here, with some egos nominating many alters, and density explosions avoided primarily by a lack of strong dependence.  By contrast, in strong friendship relations, baseline density is very low, with both reciprocity and ISP-induced closure being powerful influences.  The two show greater diminishing marginal effects, however, suggesting that either is a viable context for tie formation in strong friendships (while both must co-occur to have a comparable effect for weak friendships).  Unlike in the weak case, actors show much greater satiation for strong friendships, with each existing tie inhibiting new ties approximately 15\% of the time.  This satiation effect prevents cliques from growing without bound, despite the presence of strong closure bias.  Strong friendships thus appear to operate in a \emph{resource constrained} manner, while weak friendships are kept in check by a lack of \emph{strong formation mechanisms}.  The shading of friendship ties from one regime into the other is readily captured by the pattern of biased net parameters.

\section{Discussion}

\subsection{Use of Alternative Learning Strategies} \label{sec_altlearn}

Although we have used random forests as the tool for approximating $\mathbf{E}_f \psi | Y_{obs}$, we note that this is not essential.  (Indeed, Bayes linear statistics employs standard regression for this purpose \cite{goldstein.woof:bk:2007}, and kernel methods have likewise been used in prior work \cite{fukumizu.et.al:jmlr:2013}.)  The procedure given in Section~\ref{sec_rfprevision} can easily be used (\emph{mutatis mutandis}) with any least-squares learner - while the squared-error loss is necessary to ensure that the estimation target is the conditional expectation, other nonparametric procedures (e.g., kernelized WLS, neural networks, random feature regressors, etc.) can be employed to approximate the posterior expectation surface.  Random forests are straightforward to train and are well-suited to this problem, however, and have outperformed other learning methods in our investigations to date.  

While we do not expect that alternative learning algorithms will lead to substantial improvements in approximation quality, it is plausible that better performance could be obtained by alternative choices of summary statistics.  Intuitively (but in contrast with the traditional focus in the biased network field on structure statistics), we find that the most important statistics in our simulation study tend to be properties such as specific triad statistics (021D and 003, in particular), mean square outdegree, and mean core number, with properties such as reciprocity, transitivity, and isolate fraction being important in models for specific parameters.  A wide range of statistics do contribute to the approximation, however, suggesting that no one of the properties used here directly captures the impact of the biased net statistics on graph structure.  Further work on improved feature sets may help in extracting more information from observed graphs.

\subsection{Distinguishing Between Dichotomized and Undichotomized Effects}

In the case of the Webster network, we were not able to reliably distinguish between models based on dichotomized versus undichotomized closure effects.  We observe that this is in general a challenging problem (at least under the priors used here): the accuracy of the RF procedure in identifying dichotomized effects when evaluated on the original training simulations is only $\approx 0.66$, suggesting substantial overlap in the networks produced by both models.  This can in part be accounted for by the presence of significant prior weight on models with no closure effects (which are indeed invariant to dichotomization, and for which the distinction is both impossible and meaningless); limiting prediction to cases with closure effects results in an accuracy of approximately 0.72.  Empty graphs also provide little information, and excluding them further enhances accuracy to 0.75 - removing all extremely sparse graphs with mean degree $<1$ boosts performance to 0.87.  Finally, we note that satiation effects, by preventing runaway edge formation, also make the two cases harder to distinguish: limiting prediction to cases without such effects raises accuracy to approximately 0.77, or 0.93 when sparse cases are also excluded.  

Taken together, we conclude that there are three basic effects that contribute to the challenge of identifying dichotomized models in the general case.  First, when we cannot \emph{a priori} exclude the possibility that there are no closure effects to classify, this ``blurs'' our judgment regarding the family of generating processes (by mixing the distinguishable cases with indistinguishable cases).  Second, some graphs are too sparse to be very informative about closure effects; classification will tend fail in these cases, though performance can be much better for graphs with richer structure.  And, finally, it can be difficult to distinguish between lack of density explosion due to dichotomization and the impact of inhibitory effects.  This may suggest utility in further study to identify statistics that help discriminate between these particular mechanisms.

\section{Conclusion}

The biased network program is one of the most venerable in the social network field, and pioneered many of the theoretical and methodological developments that would go on to shape modern network analysis.  Progress within the biased net framework has involved clearing up the ambiguities and approximations present in earlier iterations, with the consequence of moving the idea of ``biased nets'' away from the tracing paradigm and towards a more local paradigm that emphasizes the idea of bias events as latent structure-forming processes.  Although we have shown here that conditional specification of biased nets leads to inconsistencies, the essential intuition behind the conditional specification can be captured with the Markovian Skvoretz-Fararo biased net process, which leads to both a well-formed family of probability distributions and a clear route to simulation.  This process is also easily extensible, as shown here by the introduction of inhibition events (which block tie formation, instead of facilitating it, and are useful for avoiding degeneracy).  Inference, however, remains a challenge.  Here, we have shown that approximate Bayesian inference using random forest prevision is a promising and general scheme for inference for SFBN processes, demonstrating its use on a large friendship network and finding a relationship between closure and tie strength of the sort proposed by \citet{fararo:sn:1983}.  Although we have limited ourselves here to basic, homogeneous mechanisms (the parent, sibling, double role, and satiation biases), we observe that our approach is equally applicable to inhomogeneous effects (such as those introduced by \citet{fararo.sunshine:bk:1964} or \citet{skvoretz.fararo:cpst:1986}).

In closing, we note that the SFBN is used here as a quasi-time generative process (in the sense of \citet{butts:jms:2023}), with its equilibrium distribution taken as a model for observed network structure.  It is interesting to consider whether the dynamics of the SFBN process may themselves be of substantive interest as models for network evolution.  This would require endowing the process with additional temporal structure, as has been done with choice models in the case of stochastic actor-oriented models \citep{snijders:sm:2001} or Gibbs-like dynamics in the case of the Longitudinal ERGMs \citep{koskinen.snijders:jspi:2007}.  Bringing explicit dynamics to the biased net family would facilitate greater contact between biased net theory and other models of social process, a unifying development of which Fararo would surely have approved \citep{fararo:st:1989}.

\bibliography{ctb}


\end{document}